# Environmental impact of 'terwatt' scale Si-photovoltaics

Satish Vitta
Department of Metallurgical Engineering and Materials Science
Indian Institute of Technology Bombay
Mumbai 400076; India.

Solar photovoltaics which converts renewable solar energy into electricity have become prominent as a measure to decarbonize electricity generation. The operational carbon footprint of this technology is significantly lower compared to electricity generation using conventional non-renewable fuels. The operationalization of photovoltaic electricity generation however results in significant emissions which are loaded upfront into the atmosphere. An analysis of the embodied energy in Si solar panels manufacturing and installation, and associated $CO_2$ emissions clearly show their global warming potential. Installation of 10 $TW_p$ capacity requires ~ 150 Exa Joules of energy and results in 7 to 13 GTons of emissions depending on the embodied energy mix. Particulate and other gaseous emissions as well as water consumption will be additional. These are significant emissions requiring development of alternate low carbon footprint manufacturing technologies and clearly shows the flip side of recent prediction of 'Terawatt' scale photovoltaics.[4]

The phenomenon of global warming is well underway with the global mean surface temperature increasing by 1.2 °C compared to the pre-industrial average value and increasing at an alarming rate.[1,2] Part of this global warming is due to the total global electricity generation which is at 26,730 TWh in the year 2018 and growing at an average rate of 3.3 %.[3] This global electricity demand has been projected to increase steadily reaching ~ 40,000 TWh by the year 2050 with a possible increase to 100,000 TWh, depending on the electricity consumption pattern changes due to increased electrification policies around the world.[4,5] In order to decarbonize electricity generation, it is being prescribed to shift to electricity generation from renewable resources. Towards this goal, as much as 25 % or more of the total electricity generation is anticipated to be generated by solar photovoltaics.[5] This will propel solar photovoltaics (PV) into 'terawatt' league with the capacity requirement anywhere between 20 TW to 50 TW by the year 2050. Currently, the installed capacity of solar PV is ~ 590 $GW_p$ and every year an additional 180 $GW_p$ capacity is being added.[6,7] According to International Renewable Energy Agency projection, the installed capacity is expected to reach 8.519 $TW_p$ by the year 2050. This corresponds to a cumulative annual growth rate of ~ 9.5 %, Figure 1 and this installed capacity falls short of the projected requirement of 20 TW to 50 TW. In order to reach the projected requirement, the installed capacity has to grow at rates of 12 % or higher as shown in Figure 1. Realization of these decarbonization goals therefore necessitates large scale capacity building at high growth rates which implies that solar panels manufacturing capabilities will also have to be increased significantly.

**Si-photovoltaics:**
Several materials such as Si, CdTe, GaAs and CIGS along with others such as dye-sensitized solar cells and perovskites have been known to have the ability to convert solar energy into electrical energy with varying degrees of efficiency. The most prominent among them however has been Si, either in monocrystalline or multicrystalline form. This is because it has been extensively studied and understood for the past several decades and is highly stable for long periods of time, typically 25 years.[8,9] Technology exists to scale up production and installation to TW levels within a short period of time. These factors make it the dominant market leader with almost 95 % of installed solar panels being made of Si. A complete replacement of Si modules is therefore not expected to take place in the foreseeable future.

A typical Si based PV module is made of several different materials – metals to polymers to ceramics: Al, Cu, Ag, Pb-containing solder, ethyl vinyl acetate (EVA), Tedlar, glass and Al-frame apart from Si which is the main element. These modules in ground-based units are typically mounted on steel and concrete structures. The typical land area required to install these modules varies from 20.235 $m^2$ per $kW_p$ to 40.47 $m^2$ per $kW_p$ depending on geographical location.[10,11] The land area requirement to install 10 $TW_p$ therefore can vary from 0.2 $MKm^2$ to 0.4 $MKm^2$ and increases depending on the actual installed capacity. It should be noted here that to derive the maximum benefit in terms of solar energy conversion, they should be installed in high solar insolence locations which are located mainly around the equator. Apart from the large high solar insolence land area requirement, creation of these solar power plants requires materials such as Si, Al, Cu, Ag, Steel, Tedlar, glass and so on which are not naturally occurring. If we just consider Si, the main ingredient, its abundant in nature in the form of $SiO_2$ (sand) and silicates, but not in the highly pure elemental form Si, which is required in the PV device. The oxide $SiO_2$ can be reduced to Si using a series of reduction and purification processes. These processes however are extremely environment intensive in nature, i.e., require sand, coal, energy and water, and the processing steps result in significant gaseous and particulate emissions. Similarly, the other materials required for putting together a Si PV module are also not naturally occurring and need extensive processing before they can be used. The environmental foot print in terms of embodied energy, i.e., energy required to process raw materials into final form and gaseous emissions, $CO_2$ associated with these processes for the different materials is given Table 1. Also given is the Herfindahl-Hirschman Index, HHI for the different elements which signifies the market concentration and their future availability based on reserves. The absolute quantity of these materials required to install different solar capacities as given in Table 1, clearly shows that the environmental impact of these materials is not insignificant and needs to be considered while assessing the 'clean' credentials of solar PV.

**Embodied energy, emissions and environmental impact:**

The embodied energy and emissions are strong functions of geographical location where the panels are manufactured as the electricity generation mix – from non-renewable and renewable resources is different in different countries. These regional differences are reflected in the total energy consumed for the manufacture of multicrystalline Si panels and the associated emissions. It has been found that ~ 3000 $MJm^{-2}$ is required in China and Europe while it is 3250 $MJm^{-2}$ in USA for the production of panels. Accordingly, the $CO_2$ emissions vary from 140 $Kgm^{-2}$ to 250 $Kgm^{-2}$ for Europe and China respectively, with the value for USA lying in between.[15] These values can be taken as upper and lower bounds to determine the actual energy requirement and emissions corresponding to different installed capacities. It can be argued that technological developments in the future can lead to reduction in the embodied energy and emissions. This is certainly true if the current energy requirement and emissions are compared to the corresponding values in the year 1990 as shown in Figure 2.[16] The historic data show that the rate of reduction has decreased substantially indicating that the technology has nearly matured leading to lowest values with no significant reduction possible in the near future. The only way these can reduce will be by a processing technology that will be 'disruptive' in nature[17] or by the discovery of a new PV material which will be as robust as Si but with far less environmental foot print and cost.

The energy required to manufacture solar panels and modules and install them on the ground, i.e., construct 'solar parks' with different installed capacity has been determined and is shown in Figure 3. Depending on the installed capacity, the embodied energy varies from 150 Exa Joules to 1200 Exa Joules for installed capacities of 10 $TW_p$ and 70 $TW_p$ respectively. The embodied energy does not vary significantly with the manufacturing location of the panels as the processes used to synthesize 7N purity Si are nearly identical. The $CO_2$ emissions on the other hand increase from ~ 50 GT to 90 GT depending on the geographical location of panels manufacturing, for an installed capacity of 70 $TW_p$, Figure 4. These are extremely large quantities of embodied energy and emissions involved with just the process of manufacturing alone. These quantities will be injected into the atmosphere over the next three decades at rates corresponding to the actual growth rate of PV installation. The energy conversion efficiency however has been known to decrease with time at 1 % every year with an initial drop of 2 % in the first year.[18] Also, end-of-life panels need to be handled and recycled, if possible, which will add to energy requirement and emissions. It should be noted here that PV do not actually 'sequester' the emissions but only lead to reduced emissions for an enhanced electrical power generation **subsequent to installation**. In effect, the absolute amount of $CO_2$ in the atmosphere would increase due to 'upfront' loading which stays in the atmosphere. This would lead to a warming phenomenon and a rise in the global average temperature. The global warming potential of emissions has been estimated from historic and geographical data and is found to be ~ 1.7 °C per Tera Tons of $CO_2(e)$.[19] Hence if one considers the global warming potential of 70 $TW_p$ panels manufacturing and installations, it can be as high as 0.16 °C. This is extremely large considering the fact that only a part of the electricity demand will be met by this technology. A recent study on the global warming potential of current food production technologies clearly shows that even if emissions from all non-food sectors are halted as of the year 2020, it will be difficult to avoid global warming to 1.5 °C by the year 2050 with the existing rate of gaseous emissions.[20] As per this scenario, all emissions including those from renewable energy technologies need to be drastically reduced or stopped and most importantly large-scale sequestration has to be implemented.

The major challenge for decarbonizing electricity generation using solar energy is 'Si'. The main advantage of Si is that it has been extensively studied and understood, both scientifically and technologically. The disadvantage is the fact that the embodied energy and emissions due to panels manufacturing are still far too large to mitigate global warming as seen in Figures 3 and 4. Therefore, for Si-based PV to become truly significant in decarbonizing electricity generation, priority should be on the manufacturing technology to reduce the energy and emissions foot print significantly. Alternate $SiO_2$ reduction and purification methods which work at low temperatures[21] and also utilizing pure form of $SiO_2$, biogenic rice husk ash, as the raw material[22] can lead to reduced environmental footprint. The other materials required for the successful implementation of tera watt scale Si PV are Al, Cu and Ag. The total amount of Al, Cu and Ag produced in the year 2019 is ~ 98 MT, 20 MT and 27,000 T respectively.[23] Installation of 10 $TW_p$ Si PV on the other hand requires 190

MT, 70 MT and 0.2 MT of Al, Cu and Ag respectively. This clearly shows that significant enhancement in production of these elements, 2X, 4X and 10X, or a potential replacement will be required. It should be noted that these elements however are used for various other applications which implies that production of these elements has to increase by orders of magnitude to cater to all the applications.

      The growing electricity demand across the world with a commitment by many nations to stay the course of emissions as per Paris agreement has catapulted solar PV electricity generation to 'tera watt' scales. Si-based solar modules are the leading candidates with almost 95 % installed capacity. This technology is unique as it is amenable to quick expansion and has very low emissions foot print **once it is operationalized**. Operationalizing this technology however is extremely energy intensive and leads to significant emissions. The present study clearly illustrates those emissions due to panels manufacturing and installation alone can be as high as 90 GT which is equivalent to increasing global average temperature by 0.16 °C. This is extremely large and clearly shows the need to reduce the emissions foot print by developing alternate Si manufacturing technologies. Another area of significant concern will be availability of elements Al, Cu and Ag. Alternatives to Cu and Ag will be especially important as these have a high HHI, which means there could be constraints on resources availability to produce these elements.

**Acknowledgements:** The author acknowledges Indian Institution of Technology Bombay for the provision of infrastructure facilities.

The author declares no competing interests and no conflict of interest. This work has not received any funding from any source.

**Table 1:** The absolute quantity of different materials required to install solar farms of specific capacity together with their embodied energy, equivalent $CO_2$ emissions, land area required and the Herfindahl-Hirschmann Index (HHI) of the elements are given. The quantity of materials is estimated as per (ref. 12) while the energy and emissions are taken from (ref. 13). The HHI for different elements is given in (ref. 14).

| Installed Capacity, TW | Land Area, MKm² | Materials, MT | | | | | | | |
|---|---|---|---|---|---|---|---|---|---|
| | | Si | Glass | Steel | Concrete | Al | Cu | Ag | Plastics |
| 10 | 0.202 | 70 | 700 | 560 | 470 | 190 | 70 | 0.2 | 60 |
| 30 | 0.607 | 210 | 2100 | 1680 | 1410 | 570 | 210 | 0.6 | 180 |
| 50 | 1.011 | 350 | 3500 | 2800 | 2350 | 950 | 350 | 1.0 | 300 |
| 70 | 1.415 | 490 | 4900 | 3920 | 3290 | 1330 | 490 | 1.4 | 420 |
| EE, MJ Kg$^{-1}$ | | 2355 | 15 | 25 | 1.0 | 28 | 16.5 | 128 | 90 |
| $CO_2$, Kg Kg$^{-1}$ | | - | 0.86 | 2.0 | 0.14 | 1.67 | 0.85 | 6.31 | 3.0 |
| HHI | | 4700 | | | | 1600 | 1600 | 1200 | |

**Figure 1.** The installed solar capacity increases with increasing cumulative annual growth rate, CAGR. In order to decarbonize electricity generation and realize significant fraction of electricity generation to be from solar PV, higher CAGR would be required. A CAGR of > 12 % will result in noticeable contribution from solar PV.

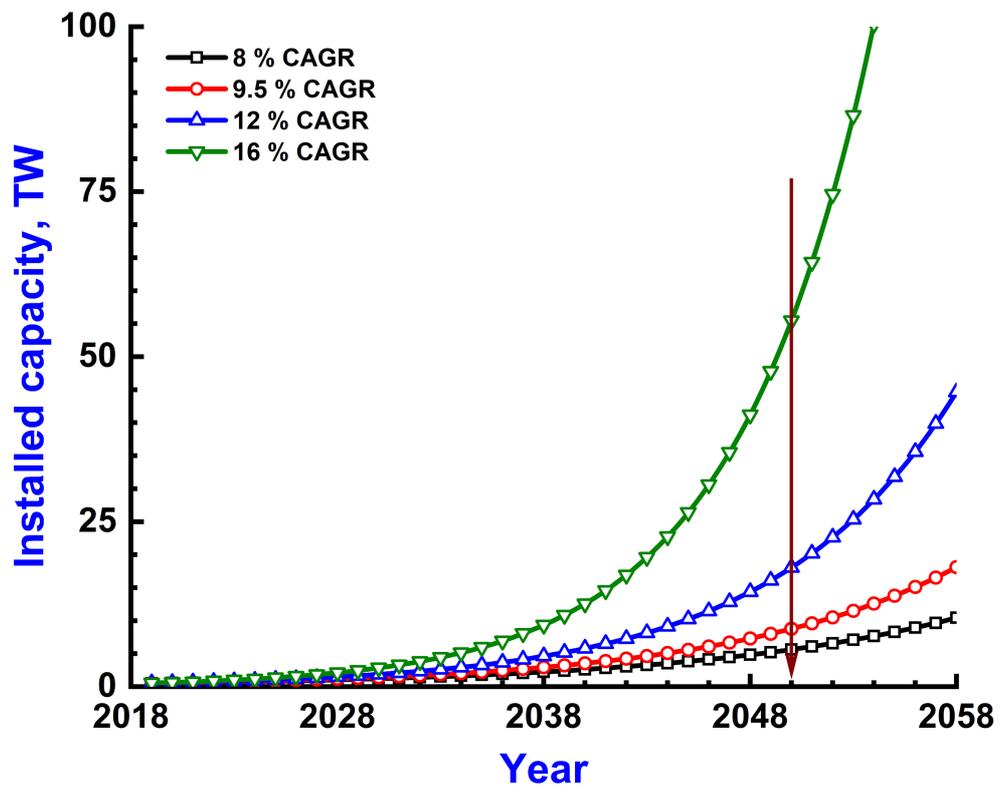

**Figure 2.** Technological developments mainly in Si processing have resulted in a continuous decrease in the embodied energy of PV panels manufacturing and the associated $CO_2$ emissions. It is however clear that the rate at which these two are decreasing has reduced indicating that further reductions may not be possible/significant.

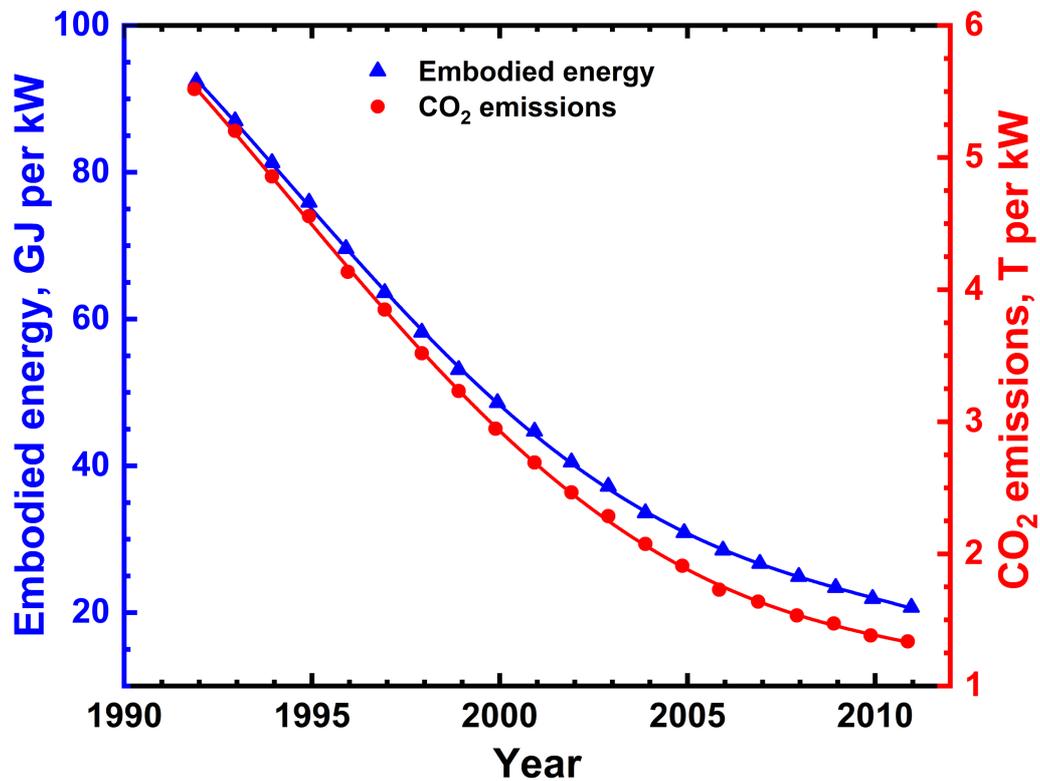

**Figure 3.** The embodied energy scales with installed capacity and can be as high as 1200 Exa Joules. The difference between manufacturing in China vis-à-vis Europe is not considerable showing that the processes are nearly identical.

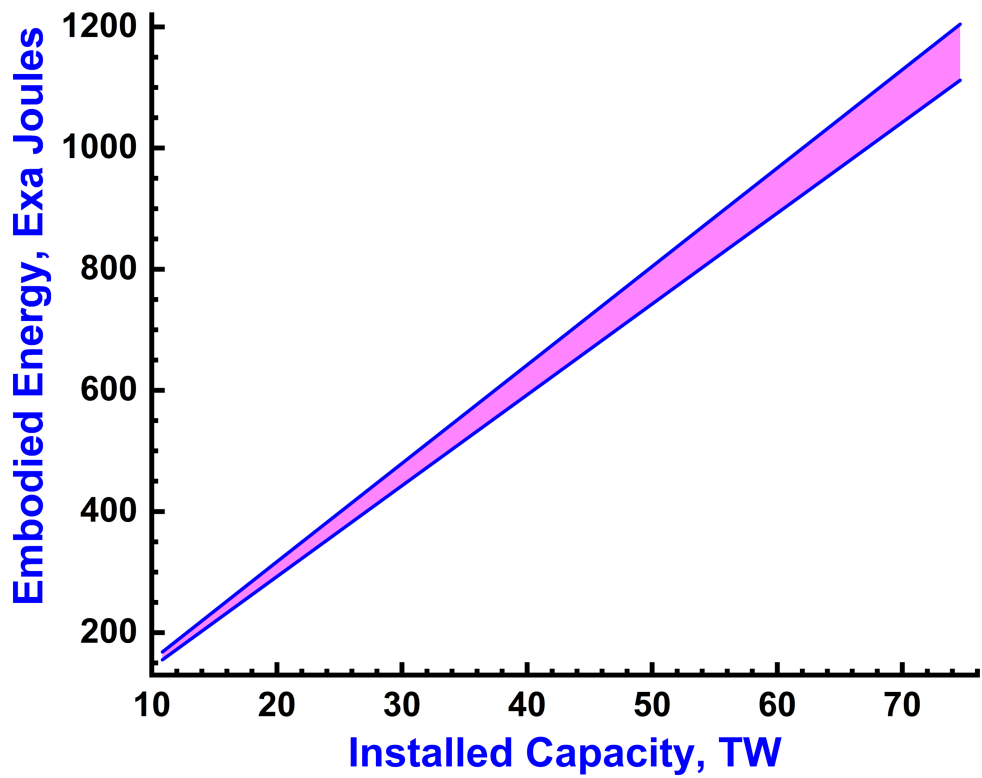

**Figure 4.** The $CO_2$ emissions scale with installed capacity and reach values of the order of 90 GT. The large difference between manufacturing in China and Europe is due to the nature of energy mix in the two places. Europe has a larger fraction of its electricity being produced from renewables compared to China.

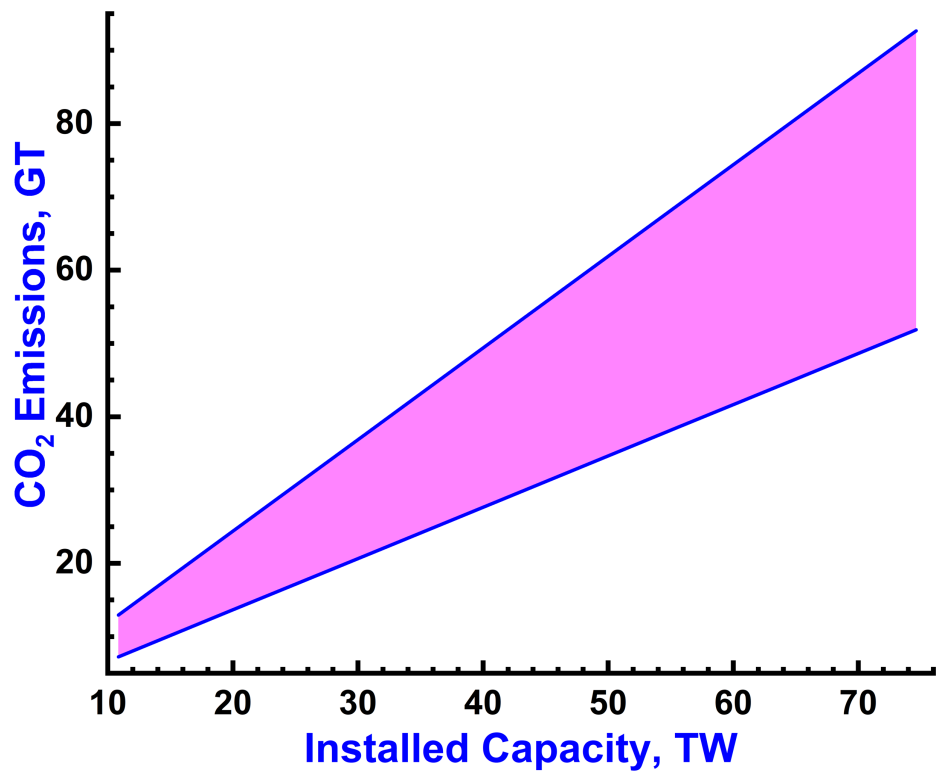